\documentstyle[12pt,fleqn]{article}
\def\mbi#1{\mbox{\boldmath$#1$}}
\def\kp{\mbi{k} \cdot \mbi{p}}

\def\beeq{\begin{equation}}
\def\eneq{\end{equation}}
\def\beeqa{\begin{eqnarray}}
\def\eneqa{\end{eqnarray}}

\begin{document}

\begin{center}

\vspace{2cm}

{\large {\bf {
Propagation of Cooper pairs 
in carbon nanotubes with superconducting correlations}}}

\vspace{1cm}

Kikuo Harigaya\footnote[1]{FAX: +81-298-61-5375; 
E-mail: \verb+k.harigaya@aist.go.jp+}

\vspace{1cm}

{\sl National Institute of Advanced Industrial
Science and Technology (AIST),\\
Umezono 1-1-1, Tsukuba 305-8568, Japan}

\end{center}

\vspace{1cm}

\noindent
{\bf Abstract}\\
Propagation of Cooper pairs in carbon nanotubes in the presence
of superconducting correlations is studied theoretically.  We
find that negative and positive currents induced by impurity
scatterings between electrons and holes cancel each other,
and the nonmagnetic impurity does not hinder the supercurrent 
in the regions where the superconducting proximity effects occur.  
The carbon nanotube is a good conductor for Cooper pairs.

\mbox{}

\noindent
Keywords:
superconductivity, impurity effects, carbon nanotubes, theory

\pagebreak

Recent investigations [1,2] show that the superconducting
proximity effect occurs when the carbon nanotubes
contact with conventional superconducting metals and wires.
The superconducting energy gap appears in the tunneling
density of states below the critical temperature $T_{\rm c}$.
On the other hand, the recent theories discuss the nature of
the exceptionally ballistic conduction [3] and the absence
of backward scattering [4] in metallic carbon nanotubes
with impurity potentials at the normal states.

In this report, we study the effects of the superconducting
pair potential on the impurity scattering processes in
metallic carbon nanotubes, using the continuum $\kp$
model for the electronic states.  We find that the carbon nanotube
is a good conductor for Cooper pairs as well as in the
normal state.  The details of this short report
have been published elsewhere [5].

We will study the metallic carbon nanotubes with the
superconducting pair potential.  The model is 
$H = H_{\rm tube} + H_{\rm pair}$.  The term
$H_{\rm tube}$ is the electronic states of the carbon
nanotubes, and the model based on the $\kp$ approximation [4]
represents electronic systems on the continuum medium: 
$H_{\rm tube} = \sum_{\mbi{k},\sigma} \Psi_{\mbi{k},\sigma}^\dagger
E_{\mbi{k}} \Psi_{\mbi{k},\sigma}$,
where $E_{\mbi{k}}$ is an energy matrix:
\beeq
E_{\mbi{k}} =
\left( \begin{array}{cccc}
0 & -i \gamma k_y & 0 & 0 \\
i \gamma k_y & 0 & 0 & 0 \\
0 & 0 & 0 & i \gamma k_y \\
0 & 0 & -i \gamma k_y & 0
\end{array} \right),
\eneq
$\mbi{k} = (0, k_y)$ is parallel with the axis of the
nanotube, and $\Psi_{\mbi{k},\sigma}$ is an
annihilation operator with four components:
$\Psi_{\mbi{k},\sigma}^\dagger =
(\psi_{\mbi{k},\sigma}^{(1)\dagger},
\psi_{\mbi{k},\sigma}^{(2)\dagger},
\psi_{\mbi{k},\sigma}^{(3)\dagger},
\psi_{\mbi{k},\sigma}^{(4)\dagger})$.
Here, the first and second elements indicate an electron at
the A and B sublattice points around the Fermi point $K$
of the graphite, respectively.  The third and fourth elements
are an electron at the A and B sublattices around the Fermi
point $K'$.  The quantity $\gamma$ is defined as $\gamma
\equiv (\sqrt{3}/2) a \gamma_0$, where $a$ is the bond
length of the graphite plane and $\gamma_0$ ($\simeq$
2.7 eV) is the resonance integral between neighboring
carbon atoms.

The second term of $H$ is the pair potential:
\beeq
H_{\rm pair} = \Delta \sum_{\mbi{k}}
(\psi_{\mbi{k},\uparrow}^{(1)\dagger}
\psi_{-\mbi{k},\downarrow}^{(1)\dagger}
+\psi_{\mbi{k},\uparrow}^{(2)\dagger}
\psi_{-\mbi{k},\downarrow}^{(2)\dagger} 
+\psi_{\mbi{k},\uparrow}^{(3)\dagger}
\psi_{-\mbi{k},\downarrow}^{(3)\dagger}
+\psi_{\mbi{k},\uparrow}^{(4)\dagger}
\psi_{-\mbi{k},\downarrow}^{(4)\dagger}
+ {\rm h.c.} )
\eneq
where $\Delta$ is the strength of the superconducting pair
correlation of an $s$-wave pairing.  We assume that the
spatial extent of the regions where the proximity effect occurs is
as long as the superconducting coherence length.

We consider the single impurity scattering
when the superconducting pair potential is present.
In the Nambu representation, the scattering $t$-matrix
at the $K$ point is
$\tilde{t}_K = \tilde{I}
[ 1 - (2/N_s) \sum_{\mbi{k}}
\tilde{G}_K (\mbi{k},\omega) \tilde{I}]^{-1}$,
where $\tilde{G}_K$ is the Nambu representation of the 
propagator of a $\pi$-electron around the Fermi point $K$ 
$G_K$ and
\beeq
\tilde{I} =
I
\left( \begin{array}{cccc}
1 & 0 & 0 & 0 \\
0 & 1 & 0 & 0 \\
0 & 0 & -1 & 0 \\
0 & 0 & 0 & -1
\end{array} \right)
\eneq
with the impurity strength $I$. 
The sign of the scattering potential for holes is
reversed from that for electrons, so the minus sign
appears at the third and fourth diagonal matrix elements.

The sum over $\mbi{k}$ is performed as in the normal
nanotubes [5], and we obtain the scattering $t$-matrix
(with the same form in the representation where 
$E=\pm \gamma k_y$ branches are diagonal):
\beeqa
\tilde{t}_K &=&
\frac{I}{1+(I \rho \pi)^2} \nonumber \\
&\times& \left( \begin{array}{cccc}
1+\alpha \omega & 0 & -\alpha \Delta & 0 \\
0 & 1+\alpha \omega & 0 & -\alpha \Delta \\
-\alpha \Delta & 0 & -1+\alpha \omega & 0 \\
0 & -\alpha \Delta & 0 & -1+\alpha \omega
\end{array} \right)
\eneqa
where $\rho$ is the density of states at the Fermi energy and
$\alpha = I \rho \pi i / \sqrt{\omega^2 - \Delta^2}$.

Hence, we find that the off-diagonal matrix elements
become zero in the diagonal $2\times2$ submatrix.  This
implies that the backward scatterings of electron-line
and hole-like quasiparticles vanish in the presence of
the proximity effects, too.  Off-diagonal $2\times2$
submatrix has the diagonal matrix elements whose
magnitudes are proportional to $\Delta$.  The finite
correlation gives rise to backward scatterings of the hole
of the wavenumber $-k_y$ when the electron with $k_y$ is incident.
The back scatterings of the electrons with the wavenumber
$-k_y$ occur for the incident holes with $k_y$, too.
Negative and positive currents induced by such the two
scattering processes cancel each other.  Therefore,
the nonmagnetic impurity {\sl does not hinder the
supercurrent} in the regions where the superconducting
proximity effects occur.  This effect is interesting
in view of the recent experimental progress of the
superconducting proximity effects [1,2].

In summary, we have investigated the effects of the
superconducting pair potential on the impurity
scattering processes in metallic carbon nanotubes.
The negative and positive currents induced by the 
impurity scatterings between electrons
and holes cancel each other.  Therefore, the carbon
nanotube is a good conductor for the Cooper pairs
coming from the proximity effects.

\pagebreak

\end{document}